# A novel device for controlling the flow of information based on Weyl fermions and some interesting remarks regarding the electromagnetic interactions of high energy particles

Georgios N. Tsigaridas[1,*], Aristides I. Kechriniotis[2], Christos A. Tsonos[2] and Konstantinos K. Delibasis[3]

[1]Department of Physics, School of Applied Mathematical and Physical Sciences, National Technical University of Athens, GR-15772 Zografou Athens, Greece

[2]Department of Physics, University of Thessaly, GR-35100 Lamia, Greece

[3]Department of Computer Science and Biomedical Informatics, University of Thessaly, GR-35131 Lamia, Greece

[*]Corresponding Author. E-mail: gtsig@mail.ntua.gr

**Abstract**

In this work we propose a novel device for controlling the flow of information using Weyl fermions. Based on a previous work of our group, we show that it is possible to fully control the flow of Weyl fermions on several different channels, by applying an electric field perpendicular to the direction of motion of the particles on each channel. In this way, we can transmit information as logical bits, depending on the existence or not of a Weyl current on each channel. We also show that the response time of this device is exceptionally low, less than 1 ps, for typical values of its parameters, allowing the control of the flow of information at extremely high rates, of the order of 100 Petabits per second. Alternatively, this device could also operate as an electric field sensor. In addition, we demonstrate that Weyl fermions can be efficiently guided through the proposed device using appropriate magnetic fields. Finally, we discuss some particularly interesting remarks regarding the electromagnetic interactions of high energy particles.

## 1. Introduction

In 1929, German physicist and mathematician Hermann Weyl predicted the existence of massless fermions that carry electric charge, named as Weyl fermions [1]. The nature of these particles suggests that they possess a high degree of mobility, moving very quickly on the surface of a crystal with no backscattering, offering substantially higher efficiency and lower heat generation compared to conventional electronics. Furthermore, these particles possess a special form of chirality with their spin being either parallel or anti-parallel to their direction of motion, referred to as positive and negative helicity, respectively. At the time of writing no such particles have been observed in nature, as free particles.

However, in 2015, an international research team led by scientists at Princeton University, detected Weyl fermions as emergent quasiparticles in synthetic crystals of the semimetal tantalum arsenide (TaAs) [2]. Independently, in the same year, a research team led by M. Soljacic at the Massachusetts Institute of Technology, also observed Weyl like excitations in photonic crystals [3]. These discoveries offer the opportunity to design and develop novel devices based on Weyl fermions instead of electrons, leading to a new branch of electronics, Weyltronics [4-10]. It should also be noted that, recently, thin films of Weyl semimetals have also been realized [4], facilitating further the development of Weyltronic devices.

In this work we describe the principles of operation of a novel device for controlling the flow of information using Weyl fermions, referred to as Weyl Parallel Switch, WPS. This device is expected to offer significant advantages over similar devices based on conventional electronics, as exceptionally low response time, increased power efficiency and extremely high bandwidth. Furthermore, due to the remarkable property of Weyl particles to be able to exist in the same quantum state in a wide variety of electromagnetic fields [11, 12], we anticipate that the proposed device will offer enhanced robustness against electromagnetic perturbations, enabling it to be used efficiently even in environments with high level of electromagnetic noise. Therefore, WPS is expected to play an important role in the emerging field of Weyltronics [4-10] and find significant applications in several fields, as telecommunications, signal processing, classical and quantum computing, etc. In addition, in section 3, we calculate the magnetic fields that could be used to fully control the transverse spatial distribution of Weyl fermions and guide them through the proposed device. Finally, in section 4, we discuss a particularly interesting remark regarding the electromagnetic interactions of high energy particles.

**2. Design and characteristics of the proposed device**

For designing the proposed device, we rely on the theory developed in a previous work of our group [11], where we have shown that it is possible to fully control the localization of Weyl particles applying an electric field perpendicular to their direction of motion. In more detail, the radius of the region where the Weyl particle is confined is given by the formula (derived in Eq. (36) of [11])

$$r(t) = \frac{r_0}{1 \pm 2qr_0|\mathbf{E}|t} \tag{1}$$

where $r_0$ is the initial value of the radius, prior to the application of the electric field, $q$ is the electric charge of the particle and $|\mathbf{E}|$ is the magnitude of the electric field. The sign in the denominator of Eq. (1) depends on the direction of the electric field relative to the angular velocity of the particles. In more detail, in the case of Weyl particles with positive helicity, the radius decreases if the electric field is antiparallel to the vector of the angular velocity of the particles and increases otherwise. The opposite is true for Weyl particles with negative helicity.

According to Eq. (1) if the radius increases with time, it becomes infinite after a time interval equal to (Eq. (37) in [11])

$$\Delta t = \frac{1}{2qr_0|\mathbf{E}|} \quad (2)$$

If the electric field continues to be applied, the radius becomes negative and decreases in magnitude, implying that the Weyl particle becomes localized again with the vector of the angular velocity pointing to the opposite direction.

It should also be noted that Eqs. (1) and (2) are expressed in natural units, where $\hbar = c = 1$. In S.I. units they take the form:

$$r(t) = \frac{r_0}{1 \pm (2qr_0|\mathbf{E}|/\hbar)t} \quad (3)$$

and

$$\Delta t = \frac{\hbar}{2qr_0|\mathbf{E}|} \quad (4)$$

respectively.

From the above analysis it becomes evident that it is possible to fully control the propagation of Weyl particles, using simple electric fields perpendicular to their direction of motion. In more detail, if we assume that a Weyl particle moves initially on a straight line and an electric field perpendicular to its direction of motion is applied at $t = 0$, then the particle becomes localized and is confined to a region of radius $r_0$, after a time interval given by Eq. (4).

This behavior can be utilized for developing a device for controlling the flow of information on multiple channels simultaneously. This device herein forth called *Weyl Parallel Switch*, WPS, is shown in the schematic diagram of figure 1.

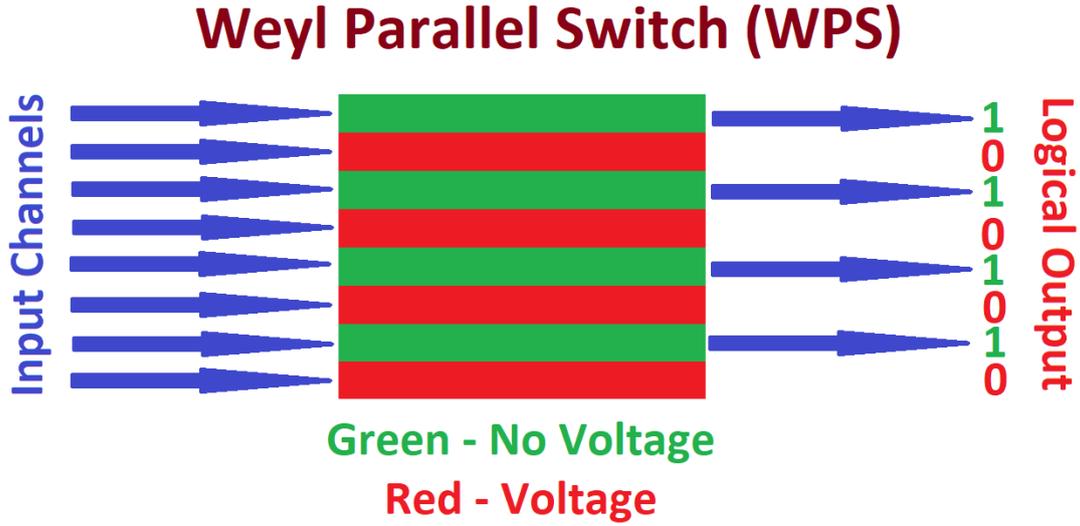

**Figure 1**: Schematic diagram of a device for controlling the flow of information based on Weyl particles.

The proposed WPS consists of a slab of a material supporting Weyl particles. An array of capacitors is constructed on this material to control the motion of Weyl fermions on each channel, by adjusting the voltage applied to the capacitor corresponding to this channel. If we assume that no voltage is applied to the capacitors, Weyl particles move along straight lines on each channel, transferring a current to the output of the channel. On the other hand, if a voltage is applied to the capacitor, the resulting electric field, which is perpendicular to the direction of motion of the particles, will confine them to a circular region of radius $r_0$. Consequently, no current will be delivered to the output of this channel.

Therefore, it is possible to control the flow of the current on each channel through the voltage applied to the capacitor corresponding to this channel. Consequently, we can control the flow of information through the channels, supposing, e.g., that the presence of a current corresponds to a logical "one" and its absence to a logical "zero", as shown in figure 1. It should also be noted that, using the proposed device we can control the flow of many bits of information simultaneously, equal to the number of channels.

The time required for the confinement of Weyl particles can be easily calculated using Eq. (4), which can also be written in the following form:

$$\frac{q r_0 \Delta V \Delta t}{d} = \frac{\hbar}{2} \qquad (5)$$

where $\Delta V$ is the voltage applied to each capacitor and $d$ is the distance between the plates of the capacitor.

As far as the width of the channel is concerned, it can be determined by the radius of the area where Weyl particles are confined, which, according to Eq. (5), is given by the formula:

$$r_0 = \frac{\hbar}{2q|\mathbf{E}|\Delta t} \tag{6}$$

Obviously, the width of the channel is equal to $2r_0$. Assuming that the charge of the particles is equal to the electron charge, it is easy to estimate the numerical value of the channel width as function of the amplitude of the electric field $|\mathbf{E}|$ and its application time $\Delta t$, given by the following formula:

$$w_{ch} = 2r_0 = \frac{\hbar}{q|\mathbf{E}|\Delta t} = \frac{6.58}{|\mathbf{E}|\Delta t} \times 10^{-16} m = \frac{0.658}{|\mathbf{E}|\Delta t} fm \tag{7}$$

which is exceptionally small. For example, if $|\mathbf{E}| = 10^4 \, V/m$ and $\Delta t = 1 \, ps$, the above formula implies that $w_{ch} = 65.8 \, nm$. Obviously, the channel width can become much smaller if the amplitude or the application time of the electric field is increased.

As far as the type of material which is preferable for the proposed device is concerned, we would like to mention that we have made no assumptions regarding the properties of the material. Consequently, any material supporting Weyl particles should be suitable for the proposed device. For practical reasons, we would prefer materials that can be shaped in the form of thin films [4], as it is also mentioned in the introduction.

Furthermore, assuming that the charge of Weyl particles is equal to the electron charge, the distance between the plates of the capacitor is $d = 1 \, mm$ and the voltage applied to each capacitor is $\Delta V = 10 \, V$, Eq. (5) implies that the time interval required to confine Weyl particles to a region of radius $r_0 = 50 \, nm$ is equal to $\Delta t = 0.658 \, ps$. Consequently, the response time of the proposed device is exceptionally low for typical values of its parameters, increasing further its efficiency. Here, it should also be mentioned that, if a voltage with opposite polarity is applied to a channel with confined particles and no current flow, then Weyl particles in this channel will become again delocalized and the current will reappear at the output of this channel. Obviously, the voltage must have the same magnitude and be applied for the same amount of time with the one used for confining Weyl particle.

Thus, it is possible to switch between logical "zeros" and "ones" - and vice versa – with response time of the order of 1 ps, for typical values of the parameters. In addition, assuming that the width of each channel is of the order of $2r_0$ and the full width of the material used in this device is of the order of 1 cm, we obtain that the device can support up to $10^5$ channels. This practically means that, using WPS, we can control the flow of information at a rate of the order of $10^{17}$ bits per second, $100 \, Pbps$, which is exceptionally difficult to achieve using conventional electronics.

Furthermore, the use of Weyl particles instead of electrons for transporting information offers higher transfer speeds, twice as fast as in graphene and up to 1000 times higher compared to conventional semiconductors [2, 10], and more efficient energy flow, substantially reducing heat generation due to collisions with the ions of the lattice. This practically means that the energy consumption of WPS, and any other device based on Weyl particles, is expected to be orders of magnitude lower than the consumption of devices based on conventional electronics.

In addition, as shown in [11, 12], Weyl particles have the remarkable property to be able to exist in the same quantum state, under a wide variety of electromagnetic fields. Specifically, as shown in [11], the quantum state of Weyl particles will not be affected by the presence of a wide variety of electromagnetic fields, which, in S.I. units, are given by the following formulae:

$$\mathbf{E}_s(\mathbf{r},t) = -\frac{1}{q}\left[\sin\theta\cos\varphi\frac{1}{c}\frac{\partial s}{\partial t} + \frac{\partial s}{\partial x} + \frac{s}{c}\left(\cos\theta\cos\varphi\frac{d\theta}{dt} - \sin\theta\sin\varphi\frac{d\varphi}{dt}\right)\right]\mathbf{i}$$
$$-\frac{1}{q}\left[\sin\theta\sin\varphi\frac{1}{c}\frac{\partial s}{\partial t} + \frac{\partial s}{\partial y} + \frac{s}{c}\left(\cos\theta\sin\varphi\frac{d\theta}{dt} + \sin\theta\cos\varphi\frac{d\varphi}{dt}\right)\right]\mathbf{j}$$
$$-\frac{1}{q}\left(\cos\theta\frac{1}{c}\frac{\partial s}{\partial t} + \frac{\partial s}{\partial z} + \frac{s}{c}\sin\theta\frac{d\theta}{dt}\right)\mathbf{k} \quad (8)$$

$$\mathbf{B}_s(\mathbf{r},t) = \frac{1}{qc}\left(-\sin\theta\sin\varphi\frac{\partial s}{\partial z} + \cos\theta\frac{\partial s}{\partial y}\right)\mathbf{i} + \frac{1}{qc}\left(\sin\theta\cos\varphi\frac{\partial s}{\partial z} - \cos\theta\frac{\partial s}{\partial x}\right)\mathbf{j}$$
$$+\frac{1}{qc}\sin\theta\left(-\cos\varphi\frac{\partial s}{\partial y} + \sin\varphi\frac{\partial s}{\partial x}\right)\mathbf{k}$$

where $\theta$, $\varphi$ are the polar and azimuthal angle, respectively, corresponding to the propagation direction of the particles, and $s$ is an arbitrary real function of the spatial coordinates and time. It should also be mentioned that, if the above electromagnetic fields are given in S.I. units, the function $s(\mathbf{r},t)$ should be expressed in Joules. As an example, we suppose that Weyl particles move at the plane $\theta = \pi/2$. Then, the electromagnetic fields given by Eq. (8) take the simplified form:

$$\mathbf{E}_s(\mathbf{r},t) = -\frac{1}{q}\left(\cos\varphi\frac{1}{c}\frac{\partial s}{\partial t} + \frac{\partial s}{\partial x} - \frac{s}{c}\sin\varphi\frac{d\varphi}{dt}\right)\mathbf{i} - \frac{1}{q}\left(\sin\varphi\frac{1}{c}\frac{\partial s}{\partial t} + \frac{\partial s}{\partial y} + \frac{s}{c}\cos\varphi\frac{d\varphi}{dt}\right)\mathbf{j}$$
$$\mathbf{B}_s(\mathbf{r},t) = -\frac{1}{qc}\sin\varphi\frac{\partial s}{\partial z}\mathbf{i} + \frac{1}{qc}\cos\varphi\frac{\partial s}{\partial z}\mathbf{j} + \frac{1}{qc}\left(-\cos\varphi\frac{\partial s}{\partial y} + \sin\varphi\frac{\partial s}{\partial x}\right)\mathbf{k} \quad (9)$$

where the azimuthal angle $\varphi$ is constant in the case of free particles, when no voltage is applied. However, in the case of particles confined by a voltage $\Delta V$, its evolution is governed by the following differential equation:

$$\frac{d^2\varphi}{dt^2} = -2q\frac{c}{\hbar}\frac{\Delta V}{d} \quad (10)$$

leading to Eq. (5) regarding the time dependence of the radius of the confined particle, as function of the applied voltage. Thus, WPS is expected to offer enhanced robustness against electromagnetic perturbations caused by the aforementioned electromagnetic fields, since the quantum state of Weyl particles will not be affected by the presence of the wide variety of electromagnetic fields given by the above formulae.

Finally, it should be mentioned that the proposed device could also operate as an electric field sensor. In more detail, the presence of an electric field, perpendicular to the Weyl current propagating in a specific channel of the device, could alter the propagation direction of the Weyl particles, interrupting the current in this channel. Specifically, Eq. (7) implies that, for a channel width $w_{ch}$ equal to 658 nm, WPS could detect an electric field of magnitude $|\mathbf{E}| = 10^{-6} V/m$ within a time interval of 1 ms. Obviously, the sensitivity of the device as an electric field sensor will improve increasing the width of the channel.

## 3. Controlling the spatial distribution of Weyl particles using appropriate magnetic fields

To guide Weyl fermions to the proposed device, we could use appropriate magnetic fields, as described below. Indeed, it is easy to verify that the spinor:

$$\psi_m = f(x,y)\begin{pmatrix}1\\0\end{pmatrix}\exp\left[iE_0(z-t)\right] \quad (11)$$

describing Weyl particles with positive helicity moving along the $+z$ direction with energy $E_0$ and transverse spatial distribution given by the arbitrary real function $f(x,y)$, is solution to the Weyl equation in the form given by Eq. (1) in [11], for the following 4-potential:

$$(a_0, a_1, a_2, a_3) = \left(0, \frac{1}{f}\frac{\partial f}{\partial y}, -\frac{1}{f}\frac{\partial f}{\partial x}, 0\right) \quad (12)$$

The electromagnetic field corresponding to the above 4-potential can be easily calculated through the formulae [14, 15]:

$$\mathbf{E} = -\nabla U - \frac{\partial \mathbf{A}}{\partial t}, \quad \mathbf{B} = \nabla \times \mathbf{A} \quad (13)$$

where $U = a_0/q$ is the electric potential and $\mathbf{A} = -(1/q)(a_1\mathbf{i} + a_2\mathbf{j} + a_3\mathbf{k})$ is the magnetic vector potential. Using Eq. (13), we obtain the electromagnetic field corresponding to the above 4-potential:

$$\mathbf{E} = 0, \quad \mathbf{B} = -\frac{1}{q}\frac{1}{f^2}\left[\left(\frac{\partial f}{\partial x}\right)^2 + \left(\frac{\partial f}{\partial y}\right)^2 - f\left(\frac{\partial^2 f}{\partial x^2} + \frac{\partial^2 f}{\partial y^2}\right)\right]\mathbf{k} \quad (14)$$

Thus, it is possible to fully control the transverse spatial distribution of Weyl particles using the magnetic field given by Eq. (14), along their propagation direction.

As an example, we consider that $f(x,y)$ is given by a generalized super-gaussian orthogonal distribution of the form:

$$f(x,y) = \exp\left[-\left(\frac{(x-x_0)^2}{\sigma_x^2}\right)^{p_x} - \left(\frac{(y-y_0)^2}{\sigma_y^2}\right)^{p_y}\right] \quad (15)$$

where $x_0, y_0$ are arbitrary real constants corresponding to the center of the distribution, and $\sigma_x, \sigma_y, p_x, p_y$ are arbitrary positive constants corresponding to the widths and the exponents of the distribution, respectively. According to Eq. (12), the 4-potential corresponding to the above distribution is the following:

$$(a_0, a_1, a_2, a_3) = \left(0, -\frac{2p_y(y-y_0)}{\sigma_y^2}\left(\frac{(y-y_0)^2}{\sigma_y^2}\right)^{-1+p_y}, \frac{2p_x(x-x_0)}{\sigma_x^2}\left(\frac{(x-x_0)^2}{\sigma_x^2}\right)^{-1+p_x}, 0\right)$$

(16)

and the magnetic field given by Eq. (14) becomes:

$$\mathbf{B} = -\frac{2}{q}\left[\frac{p_x(-1+2p_x)}{\sigma_x^2}\left(\frac{(x-x_0)^2}{\sigma_x^2}\right)^{-1+p_x} + \frac{p_y(-1+2p_y)}{\sigma_y^2}\left(\frac{(y-y_0)^2}{\sigma_y^2}\right)^{-1+p_y}\right]\mathbf{k} \quad (17)$$

Similarly, in the case of Weyl fermions with negative helicity, described by Eq. (2) in [11], the spinor corresponding to particles moving along the $+z$ direction with energy $E_0$ and transverse spatial distribution given by the function $f(x,y)$ is the following:

$$\psi'_m = f(x,y)\begin{pmatrix}0\\1\end{pmatrix}\exp[iE_0(z-t)] \quad (18)$$

The 4-potential given by Eq. (10) becomes:

$$(a'_0, a'_1, a'_2, a'_3) = \left(0, -\frac{1}{f}\frac{\partial f}{\partial y}, \frac{1}{f}\frac{\partial f}{\partial x}, 0\right) = -(a_0, a_1, a_2, a_3) \quad (19)$$

Furthermore, according to Eq. (13), the electromagnetic field corresponding to the above 4-potential takes the form:

$$\mathbf{E}' = \mathbf{0}, \quad \mathbf{B}' = \frac{1}{q}\frac{1}{f^2}\left[\left(\frac{\partial f}{\partial x}\right)^2 + \left(\frac{\partial f}{\partial y}\right)^2 - f\left(\frac{\partial^2 f}{\partial x^2} + \frac{\partial^2 f}{\partial y^2}\right)\right]\mathbf{k} = -\mathbf{B} \quad (20)$$

Thus, the electromagnetic 4-potential and field required to manipulate the transverse spatial distribution of Weyl particles with negative helicity is opposite to the one required to control the spatial distribution of particles with positive helicity.

Finally, a particularly important remark is that, according to Theorem 3.1 in [12], the spinors given by Eq. (11) will also be solutions of the Weyl equation (1) in [11] for an infinite number of 4-potentials, given by the formula:

$$b_\mu = a_\mu + \kappa_\mu s(\mathbf{r},t), \quad \mu = 0,1,2,3 \tag{21}$$

where

$$(\kappa_0, \kappa_1, \kappa_2, \kappa_3) = \left(1, -\frac{\psi^\dagger \sigma^1 \psi}{\psi^\dagger \psi}, -\frac{\psi^\dagger \sigma^2 \psi}{\psi^\dagger \psi}, -\frac{\psi^\dagger \sigma^3 \psi}{\psi^\dagger \psi}\right) = (1,0,0,-1) \tag{22}$$

and $s(\mathbf{r},t)$ is an arbitrary real function of the spatial coordinates and time.

Similarly, in the case of particles with negative helicity, the spinors given by Eq. (18) will also be solutions to the Weyl equation (2) in [11], for the following 4-potentials:

$$b'_\mu = a'_\mu + \kappa'_\mu s(\mathbf{r},t), \quad \mu = 0,1,2,3 \tag{23}$$

where

$$(\kappa'_0, \kappa'_1, \kappa'_2, \kappa'_3) = \left(1, \frac{\psi'^\dagger \sigma^1 \psi'}{\psi'^\dagger \psi'}, \frac{\psi'^\dagger \sigma^2 \psi'}{\psi'^\dagger \psi'}, \frac{\psi'^\dagger \sigma^3 \psi'}{\psi'^\dagger \psi'}\right) = (1,0,0,-1) = (\kappa_0, \kappa_1, \kappa_2, \kappa_3) \tag{24}$$

According to Eq. (13), the 4-potentials:

$$b_\mu - a_\mu = b'_\mu - a'_\mu = (1,0,0,-1) s(\mathbf{r},t) \tag{25}$$

correspond to the following electromagnetic fields:

$$\mathbf{E}_s(\mathbf{r},t) = -\frac{1}{q}\left(\frac{\partial s}{\partial x}\mathbf{i} + \frac{\partial s}{\partial y}\mathbf{j} + \left(\frac{\partial s}{\partial t} + \frac{\partial s}{\partial z}\right)\mathbf{k}\right),$$
$$\mathbf{B}_s(\mathbf{r},t) = \frac{1}{q}\left(\frac{\partial s}{\partial y}\mathbf{i} - \frac{\partial s}{\partial x}\mathbf{j}\right) \tag{26}$$

This suggests that the state of Weyl particles will not be affected if any of the above electromagnetic fields is added to the magnetic fields given by Eqs. (14), (20), corresponding to particles with positive and negative helicity, respectively. Thus, the process of controlling the spatial distribution of Weyl particles through appropriate magnetic fields is robust against electromagnetic perturbations, at least of the form described by Eq. (26).

**4. On the electromagnetic interactions of high energy particles**

As mentioned in our first work on this subject [12], a spinor describing a free particle is degenerate if it satisfies the following condition in natural units:

$$p = E + m \tag{27}$$

where $p$ is the modulus of the momentum of the particle, $m$ is its mass and $E$ its total energy. In SI units the above condition becomes:

$$pc = E + mc^2 \tag{28}$$

Using the well-known identity of the special theory of relativity, $E^2 = p^2c^2 + m^2c^4$, Eq. (28) takes the following form:

$$pc = \sqrt{p^2c^2 + m^2c^4} + mc^2 \tag{29}$$

which is approximately satisfied if $pc \gg mc^2$. This practically means that the total energy of the particle should be much greater than its rest energy. Thus, the theory of degeneracy, as described in [12], is also expected to be approximately valid for high energy particles. Furthermore, the higher the total energy of the particle compared to its rest energy, the more evident the effects of degeneracy are expected to be, as it is also discussed in [16], in section 4. In this case, the particle will be mostly unaffected by the wide variety of electromagnetic fields described by Eq. (8) setting $d\theta/dt = d\varphi/dt = 0$.

For example, as discussed in [16], the state of the particle will be mostly unaffected by the presence of an electric field, of arbitrary time-dependence, applied along the propagation direction of the particle, or an electromagnetic wave, of arbitrary polarization, propagating along the direction of motion of the particle.

Supposing that the particle under consideration is a proton, having rest energy 938 MeV, the effects of degeneracy are expected to be evident for values of the total energy above 10 GeV, which is far below the energy obtained in modern particle accelerators. For example, at the Large Hadron Collider (LHC) at CERN, the energy of the protons can reach values up to 6.5 TeV. For these values of the energy of the protons the effects of degeneracy are expected to be prominent, which may lead to major breakthroughs regarding the analysis of the experiments performed at LHC and other particle accelerating facilities.

Furthermore, based on similar arguments, the theory of degeneracy is also expected to be applicable to high-energy cosmic rays, and may help answer some open questions especially regarding ultra-high energy cosmic rays. For example, the energy of the Oh-My God and Amaterasu particles, seems to exceed the Greisen–Zatsepin–Kuzmin (GZK) limit. This practically means that these particles interact less than expected with the microwave background radiation (CMB), which could be explained using the theory of degeneracy [12, 16], predicting that a high-energy particle does not interact significantly with an electromagnetic field propagating along its direction of motion.

Another interesting remark is that, according to Eq. (2) in [17], the arrival time of photons from distant sources, e.g. gamma-ray bursts, is expected to depend on the energy of the photons. In more detail, higher energy photons are expected to interact more strongly with the virtual particles of the quantum vacuum, resulting to longer arrival times (lower velocities). This is also consistent with our theory, since the virtual particles seem less energetic to higher energy photons, suppressing the effects of degeneracy, and increasing the interactions between the photons and the virtual particles, resulting to a reduction in their velocity and longer arrival time. On the other hand, virtual particles seem more energetic to lower frequency photons, making more prominent the effects of degeneracy, resulting to suppressed interactions between the photons and the virtual particles. Thus, the velocity of lower energy photons is expected to be less affected by the interactions with the virtual particles of the quantum vacuum, in agreement with equations (1) and (2) in [17].

## 5. Conclusions

In conclusion, we have described the principles of operation and the main properties of a simple and efficient device, the Weyl Parallel Switch (WPS), for controlling the flow of information, based on Weyl fermions. This device has the advantage that it can control the flow of information along multiple channels simultaneously. In addition, the response time is exceptionally low, under 1 ps for typical values of the parameters, enabling the control of information flow at a rate of the order of 100 Petabits per second for a device with dimensions of the order of 1 cm, which is exceptional difficult to achieve using conventional electronics. Furthermore, the remarkable property of Weyl particles to be able to exist in the same quantum state under a wide variety of electromagnetic fields [11, 12], provides enhanced robustness against electromagnetic perturbations, offering the opportunity to use the device in environments with high level of electromagnetic noise. Consequently, WPS is expected to play an important role in the emerging field of Weyltronics [4-10], and could be utilized in a variety of applications, as telecommunications, signal processing, classical and quantum computing, etc. In addition, WPS could also operate as a sensitive electric field sensor. Furthermore, we have proposed a method to fully control the transverse spatial distribution of Weyl fermions using appropriate magnetic fields, which could be used to guide Weyl fermions through the proposed device. Finally, we have discussed some very interesting remarks regarding the electromagnetic interactions of high energy particles, where we have shown that the effects of degeneracy could also be applicable in this case.

## References

[1] Hermann Weyl, Gravitation and the electron, Proc. Natl. Acad. Sci. U.S.A. **15**, 323-334 (1929) DOI: 10.1073/pnas.15.4.323

[1] Hermann Weyl, Gravitation and the electron, Proc. Natl. Acad. Sci. U.S.A. **15**, 323-334 (1929) DOI: 10.1073/pnas.15.4.323

[1] Hermann Weyl, Gravitation and the electron, Proc. Natl. Acad. Sci. U.S.A. **15**, 323-334 (1929) DOI: 10.1073/pnas.15.4.323

[1] Hermann Weyl, Gravitation and the electron, Proc. Natl. Acad. Sci. U.S.A. **15**, 323-334 (1929) DOI: 10.1073/pnas.15.4.323

[2] Su-Yang Xu et al., Discovery of a Weyl fermion semimetal and topological Fermi arcs, arXiv:1502.03807 [cond-mat.mes-hall], Science 349, 613-617 (2015) DOI:10.1126/science.aaa9297

[3] Ling Lu et al., Experimental observation of Weyl points, arXiv:1502.03438 [cond-mat.mtrl-sci], Science 349, 622-624 (2015) DOI:10.1126/science.aaa9273

[4] A. Bedoya-Pinto et al., Towards Weyltronics: Realization of epitaxial NbP and TaP Weyl Semimetal thin films, arXiv:1909.12707 [cond-mat.mes-hall], ACS Nano **14**, 4405–4413 (2020) DOI: 10.1021/acsnano.9b09997

[5] S. Singh, A. C. Garcia-Castro, I. Valencia-Jaime, F. Muñoz and A. H. Romero, Prediction of a controllable Weyl semi-metallic phase in inversion-asymmetric BiSb, arXiv:1512.00863 [cond-mat.mtrl-sci], Phys. Rev. B **94**, 161116(R) (2016) DOI: 10.1103/PhysRevB.94.161116

[6] A. C. Niemann et al., Chiral magnetoresistance in the Weyl semimetal NbP, arXiv:1610.01413 [cond-mat.mtrl-sci], Sci Rep **7**, 43394 (2017) DOI: 10.1038/srep43394

[7] J. Shen et al., 33% Giant Anomalous Hall Current Driven by both Intrinsic and Extrinsic Contributions in Magnetic Weyl Semimetal $Co_3Sn_2S_2$, arXiv:2007.08268 [cond-mat.mtrl-sci], Adv. Funct. Mater. **30,** 2000830 (2020) DOI: 10.1002/adfm.202000830

[8] Z. -X. Li, X. S. Wang, L. Song, Y. Cao, and P. Yan, Type-II Weyl Excitation in Vortex Arrays, arXiv:2109.07083 [cond-mat.mes-hall], Phys. Rev. Applied **17**, 024054 (2022) DOI: 10.1103/PhysRevApplied.17.024054

[9] N. Aryal, X. Jin, Q. Li, M. Liu, A. M. Tsvelik, and W. Yin, Robust and tunable Weyl phases by coherent infrared phonons in $ZrTe_5$, arXiv:2110.07076 [cond-mat.mtrl-sci], npj Comput Mater **8**, 113 (2022) DOI: 10.1038/s41524-022-00800-z

[10] N.P. Armitage, E.J. Mele, and A. Vishwanath, Weyl and Dirac Semimetals in three-dimensional solids, arXiv:1705.01111 [cond-mat.str-el], Rev. Mod. Phys. 90, 015001 (2018) DOI: 10.1103/RevModPhys.90.015001

[11] G. N. Tsigaridas, A. I. Kechriniotis, C. A. Tsonos and K. K. Delibasis, On the localization properties of Weyl particles, arXiv:2205.11251 [quant-ph], Ann. Phys. (Berlin) **534**, 2200437 (2022) DOI: 10.1002/andp.202200437

[12] A. I. Kechriniotis, C. A. Tsonos, K. K. Delibasis and G. N. Tsigaridas, On the connection between the solutions to the Dirac and Weyl equations and the corresponding electromagnetic 4-potentials, arXiv:1208.2546 [math-ph], Commun. Theor. Phys. **72** (2020) 045201, DOI: 10.1088/1572-9494/ab690e

[13] M. Thomson, Modern Particle Physics, Cambridge University Press, Cambridge (2013) ISBN: 9781139525367

[14] D. J. Griffiths, Introduction to Electrodynamics (4th ed.), Cambridge University Press, Cambridge (2017) ISBN: 9781108420419

[15] J. D. Jackson, Classical Electrodynamics (3rd ed.), Wiley, New York (1998) ISBN: 978-0-471-30932-1

[16] G. N. Tsigaridas, A. I. Kechriniotis, C. A. Tsonos and K. K. Delibasis, Degenerate solutions to the massless Dirac and Weyl equations and a proposed method for controlling the quantum state of Weyl particles, arXiv:2010.09846 [quant-ph], Chin. J. Phys. **77** (2022) 2324-2332, DOI: 10.1016/j.cjph.2022.04.010

[17] G. Amelino-Camelia, J. Ellis, N. E. Mavromatos, D. V. Nanopoulos and S. Sarkar, Tests of quantum gravity from observations of γ-ray bursts, Nature **393** (1998) 763-765, DOI: 10.1038/26793